\documentstyle[emulateapj,epsf]{article}

\lefthead{BAUMGARTE \& SHAPIRO}
\righthead{LUMINOSITY VERSUS ROTATION IN A SUPERMASSIVE STAR}
\begin{document}

\title{LUMINOSITY VERSUS ROTATION IN A SUPERMASSIVE STAR}

\author{Thomas W.~Baumgarte and 
        Stuart L.~Shapiro\altaffilmark{1}}
\affil{Department of Physics, University of Illinois at
        Urbana-Champaign, Urbana, Il~61801}
\altaffiltext{1}{Department of Astronomy and
        National Center for Supercomputing Applications, 
        University of Illinois at Urbana-Champaign, Urbana, Il~61801}

\begin{abstract}
We determine the effect of rotation on the luminosity of supermassive
stars.  We apply the Roche model to calculate analytically the emitted
radiation from a uniformly rotating, radiation-dominated supermassive
configuration.  We find that the luminosity at maximum rotation, when
mass at the equator orbits at the Kepler period, is reduced by $\sim
36\%$ below the usual Eddington luminosity from the corresponding
nonrotating star.  A supermassive star is believed to evolve in a
quasistationary manner along such a maximally rotating
``mass-shedding'' sequence before reaching the point of dynamical
instability; hence this reduced luminosity determines the evolutionary
timescale.  Our result therefore implies that the lifetime of a
supermassive star prior to dynamical collapse is $\sim 56\%$ longer
than the value typically estimated by employing the usual Eddington
luminosity.
\end{abstract}

\section{INTRODUCTION}

Recent observations provide strong evidence that supermassive black
holes (SMBHs) exist and are the sources that power active galactic
nuclei and quasars (see, e.g., Rees 1998, for a review and
references).  However, the scenario by which SMBHs form is still very
uncertain (see, e.g., Rees 1984, for an overview).  One promising
route is the collapse of a supermassive star (SMS).  Once they form out
of primordial gas, sufficiently massive stars will evolve
in a quasistationary manner via radiative cooling, slowly contracting until
reaching the point of onset of relativistic radial instability. At
this point, such stars undergo catastrophic collapse on a dynamical
timescale, possibly leading to the formation of a SMBH
(Bisnovatyi-Kogan, Zel'dovich \& Novikov 1966; Zel'dovich \& Novikov
1971; Shapiro \& Teukolsky 1983).

Because most objects formed in nature have some angular momentum,
rotation is likely to play a significant role in the quasistationary
evolution, as well as the final collapse of a SMS.  The slow
contraction of even a slowly rotating SMS will likely spin it up to
the mass-shedding limit, because such stars are so centrally
condensed. At the mass-shedding limit, matter on the equator moves in
a Keplerian orbit about the star, supported against gravity by
centrifugal force and not by an outward pressure gradient. The SMS
evolves in a quasistationary manner along the mass-shedding curve,
simultaneously emitting radiation, matter and angular momentum until
reaching the onset of radial instability.

In this paper, we derive the luminosity of a uniformly rotating SMS as
a function of its spin rate, up to the mass-shedding limit.  The
magnitude of the luminosity is crucial because it determines the
evolutionary timescale of the star as it evolves.  
Elsewhere we use the result to follow the slow contraction of a
cooling, rotating SMS to the onset of dynamical instability (Baumgarte
\& Shapiro 1999, hereafter Paper II). Here, however, we focus on
the emitted flux and total integrated luminosity from a stationary SMS
as a function of its rate of rotation.  [For an overview of previous
work on SMSs and references, see, e.g., Zel'dovich \& Novikov 1971;
Shapiro \& Teukolsky 1983; and Paper II.]

Even though our calculation, which is analytic up to quadrature, is
rather simple and straight-forward, we have not been able to find a
similar argument in the literature.  Previous analytical arguments
have dealt with more general rotation laws, but they adopt the slow
rotation approximation and emphasize gas-pressure atmospheres (see,
e.g., Kippenhahn 1977).  While detailed numerical calculations of
luminosities of rotating stars have been carried out for select main
sequence and massive stars (see, eg, Tassoul 1978, Table 12.1, and
references therein; Langer \& Heger 1997), we cannot find a
calculation for a SMS.  Hence independent of its relevance to the
evolution of SMSs prior to catastrophic collapse, our result may be of
interest to stellar modeling of rapidly rotating stars in the limit of
very high mass, where our calculation is applicable.

Our paper is organized as follows: in Section 2 we enumerate and
justify our assumptions. In Section 3 we briefly review the Roche
approximation, which we use to describe the outer layers of a rotating
SMS. In Section 4 we derive the flux and luminosity from the star.  In
Section 5 we briefly summarize our results and compare them with
previous calculations of rotating main sequence stars.

\section{BASIC ASSUMPTIONS}

Our analysis relies on several explicit assumptions, all of which we expect
to hold to high accuracy in SMSs.  In particular, we assume that the star is

\begin{enumerate}

\item dominated by thermal radiation pressure;

\item fully convective;

\item uniformly rotating;

\item characterized by a Rosseland mean opacity that is independent of density;

\item governed by Newtonian gravitation;

\item described by the Roche model in the outer envelope.

\end{enumerate}

For large masses, the ratio between radiation pressure, $P_r$, and gas
pressure, $P_g$, satisfies
\begin{equation} \label{beta}
\beta \equiv \frac{P_g}{P_r} = 8.49 
	\left( \frac{M}{M_{\odot}} \right)^{-1/2}
\end{equation}
(see, e.g., Shapiro \& Teukolsky 1983, eqs.~(17.2.8) and~(17.3.5));
here the coefficient has been evaluated for a composition of pure
ionized hydrogen.  For stars with $M \gtrsim 10^4 M_{\odot}$, we can
therefore neglect the pressure contributions of the plasma in
determining the equilibrium profile, even though the plasma may be
important for determining the stability of the star (Zel'dovich \&
Novikov 1971; and Shapiro \& Teukolsky 1983).  A simple proof that
SMSs are convective in this limit is given in Loeb \& Rasio (1994).
This result implies that the photon entropy per baryon,
\begin{equation}
s_r = \frac{4}{3} \,\frac{a T^3}{n_B} 
\end{equation}
is constant throughout the star, and so therefore is $\beta \approx 8
(s_r/k)^{-1}$.  Here $a$ is the radiation density constant, $n_B$ is the
baryon density, and $k$ is Boltzmann's constant.  As a consequence,
the equation of state of a SMS is that of an $n=3$ polytrope:
\begin{equation}
P = K \rho^{4/3},  \mbox{~~}
K = \left[ \Big( \frac{k}{\bar \mu m} \Big)^4 \frac{3}{a}
\frac{(1 + \beta)^3}{\beta^4} \right]^{1/3} = const,
\end{equation}
where $P$ is the pressure, $\rho$ the mass density, $m$ the atomic
mass unit and $\bar \mu$ the mean molecular weight (cf.~Clayton 1983,
eq.~2-289; note that Clayton adopts a different definition of $\beta$,
which is related to ours by $\beta_{\rm Clayton} = \beta /(1+\beta)$).

The third assumption, that the star is uniformly rotating, is probably
the most uncertain of our assumptions.  Nevertheless, it has been
argued that convection and magnetic fields provide an effective
turbulent viscosity which dampens differential rotation and brings the
star into uniform rotation (Bisnovatyi-Kogan, Zel'dovich \& Novikov
1967; Wagoner 1969).

In the high temperature, low density, strongly ionized plasma of a
SMS, Thomson scattering off free electrons is the dominant source of
opacity.  This opacity is independent of density and justifies our
fourth assumption.

We assume that gravitational fields are sufficiently weak so that we
can apply Newtonian gravity.  SMSs of interest here have $R/M \gtrsim
400$ (see Paper II), so that this assumption certainly holds.
Relativistic corrections are important for the stability of SMSs, but
can be neglected in the analysis of the equilibrium state.

Finally, the Roche approximation provides a very accurate description
of the envelope of a rotating stellar model with a soft equation of
state, as in the case of an $n=3$ polytrope (for numerical
demonstrations, see, eg, Papaloizou \& Whelan 1973 and Paper II).
Since our analysis is based on this approximation, we will briefly
review it together with some of its predictions in the following
section.

In applications to SMSs our analysis neglects electron-positron pairs
and Klein-Nishina corrections to the electron scattering opacity, 
which is valid for $M \gtrsim 10^5 M_{\odot}$ (see, e.g.,
Fuller, Woosley \& Weaver, 1986).

\section{REVIEW OF THE ROCHE MODEL}

Stars with soft equations of state are extremely centrally condensed:
they have an extended, low density envelope, while the bulk at the
mass is concentrated in the core.  For an $n=3$ polytrope, for
example, the ratio between central density to average density is
$\rho_c/ \bar \rho = 54.2$.  The gravitational force in the envelope
is therefore dominated by the massive core, and it is thus legitimate
to neglect the self-gravity of the envelope.  In the equation of
hydrostatic equilibrium,
\begin{equation} \label{hydro}
\frac{{\bf \nabla} P}{\rho} = - {\bf \nabla} (\Phi + \Phi_c),
\end{equation}
this neglect amounts to approximating the Newtonian potential $\Phi$ by
\begin{equation}
\Phi = - \frac{M}{r}
\end{equation}
(here we adopt gravitational units by setting $G \equiv 1$).
In~(\ref{hydro}) we introduce the centrifugal potential $\Phi_c$,
which, for constant angular velocity $\Omega$ about the $z$-axis, can
be written
\begin{equation} \label{centrif}
\Phi_c = - \frac{1}{2} \Omega^2\,(x^2 + y^2) = 
	- \frac{1}{2} \Omega^2 r^2 \sin^2 \theta.
\end{equation}
Integrating eq.~(\ref{hydro}) yields the Bernoulli integral
\begin{equation} \label{bernoulli}
h + \Phi + \Phi_c = H,
\end{equation}
where $H$ is a constant of integration and 
\begin{equation}
h = \int \frac{dP}{\rho} = (n+1)\,\frac{P}{\rho}
\end{equation}
is the enthalpy per unit mass.  Evaluating eq.~(\ref{bernoulli})
at the pole yields
\begin{equation}
H = - \frac{M}{R_p},
\end{equation}
since $h = 0$ on the surface of the star and $\Phi_c = 0$ along the
axis of rotation.  In the following we assume that the polar radius
$R_p$ of a rotating star is always the same as in the nonrotating case. 
This assumption has been shown numerically to be very accurate (e.g.,
Papaloizou \& Whelan 1973).

A rotating star reaches mass shedding when the equator orbits with the
Kepler frequency.  Using eqs.~(\ref{centrif}) and~(\ref{bernoulli}), it 
is easy to show
that at this point the ratio between equatorial and polar radius is
\begin{equation}
\left( \frac{R_e}{R_p} \right)_{\rm shedd} = \frac{3}{2}.
\end{equation}
The corresponding maximum orbital velocity is
\begin{equation}
\Omega_{\rm shedd} = \left( \frac{2}{3} \right)^{3/2}
	\left( \frac{M}{R_p^3} \right)^{1/2}
\end{equation}
(Zel'dovich \& Novikov 1971; Shapiro \& Teukolsky 1983).

\section{LUMINOSITY OF ROTATING STARS}


\begin{figure*} \label{fig1}
\epsscale{.4}
\plotone{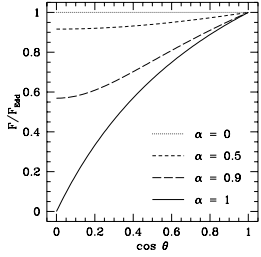}
\caption{The flux $F$ as a function of polar angle
for different values of $\alpha = \Omega /\Omega_{\rm shedd}$.  Note
that at the mass-shedding limit $\alpha = 1$ the flux vanishes on the
equator.}
\end{figure*}

According to our assumptions, the pressure in supermassive stars
is dominated by radiation pressure
\begin{equation} 
P \approx P_r = \frac{1}{3} a T^4.
\end{equation}
In the diffusion approximation, the radiation flux is everywhere given
by
\begin{equation} \label{flux}
{\bf F} = - \frac{1}{3 \kappa \rho} {\bf \nabla} U.
\end{equation}
Here $U$ the energy density of the radiation,
\begin{equation} \label{U}
U = a T^4 = 3 P,
\end{equation}
and $\kappa$ is the opacity (which we assume to be dominated
by electron scattering, $\kappa = \kappa_{\rm es}$).  
Inserting~(\ref{flux}) and~(\ref{U}) into the equation of
hydrostatic equilibrium~(\ref{hydro}) yields
\begin{equation} \label{flux1}
\kappa {\bf F} = {\bf \nabla} (\Phi + \Phi_c).
\end{equation}
In polar coordinates in an orthonormal basis, the magnitude of the
flux is
\begin{equation}
F = \left( F_{\hat r}^2 +  F_{\hat \theta}^2 \right)^{1/2}.
\end{equation}
Evaluating the gradients of $\Phi$ and $\Phi_c$ in the envelope yields
\begin{equation} \label{flux2}
F = \frac{M}{\kappa r^2} \left[
	1 - 2 \frac{\Omega^2 \sin^2 \theta}{M/r^3} +
	\left( \frac{\Omega^2 \sin \theta}{M/r^3} \right)^2 \right]^{1/2}.
\end{equation}
Introducing the dimensionless spin and radius parameters
\begin{equation}
\alpha \equiv \frac{\Omega}{\Omega_{\rm shedd}}
\end{equation}
and 
\begin{equation}
z \equiv \frac{r}{R_p},
\end{equation}
and denoting
\begin{equation} \label{f_eddington}
F_{\rm Edd} = \frac{M}{\kappa r^2},
\end{equation}
the usual Eddington flux from a spherical star, we can 
rewrite equation~(\ref{flux2}) as
\begin{equation} \label{flux3}
\frac{F}{F_{\rm Edd}} =  \Big[ 1 - 2 \big(\frac{2}{3}\big)^3
	\alpha^2 z^3 \sin^2 \theta + \big(\frac{2}{3}\big)^6
	\alpha^4 z^6 \sin^2 \theta \Big]^{1/2}.
\end{equation}

Note that, from eq.~(\ref{bernoulli}), the surface of the star is defined by
\begin{equation} \label{surf}
\Phi + \Phi_c - H = 0,
\end{equation}
or equivalently\footnote{Expanding eqs.~(\ref{flux2}) and (\ref{z}) to
lowest order in $\alpha^2$ shows that they are in perfect agreement
with eqs.~(39) and (30) in Kippenhahn (1977) for uniform rotation.
Kippenhahn's treatment allows for non-uniform rotation, but is
restricted to slow rotation.}
\begin{equation} \label{z}
\frac{4}{27} \alpha^2 z^3 \sin^2 \theta - z + 1 = 0.
\end{equation}
Given $\alpha$ and $\theta$, the value of $z$ on the surface can be
found by solving this cubic equation.  Eqs.~(\ref{flux1}) and
~(\ref{surf}) immediately imply that the flux is normal to the surface
of the star.

Evaluating eq.~(\ref{flux3}) at the surface, we plot the emergent flux
$F$ as a function of $\theta$ for different values of $\alpha$ in
Figure~1.  Note that at the mass-shedding limit, when $\alpha = 1$,
the flux vanishes at the equator (where $z = 3/2$ and $\sin \theta =
1$).  This, of course, is an immediate consequence of hydrostatic
equilibrium: at mass shedding, the centrifugal force exactly balances
the gravitational force at the equator, so that the pressure gradient
vanishes (eq.~(\ref{hydro})).  For radiation-dominated stars,
eq.~(\ref{flux1}) then implies that the flux has to vanish.

\begin{figure*} \label{fig2}
\epsscale{.4}
\plotone{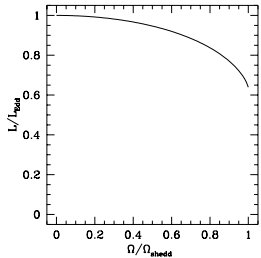}
\caption{The luminosity $L$ as a function of the orbital velocity
$\Omega$.  At the mass shedding limit, the luminosity is reduced by
36 \%.}
\end{figure*}

The total luminosity can be found by integrating
\begin{equation}
L = \int_{\cal A} {\bf F \cdot d {\cal A}} = 
	\int_{\cal A} F d {\cal A}
\end{equation}
over the surface ${\cal A}$ of the star.  The surface element 
$d {\cal A}$ can be written
\begin{eqnarray}
d {\cal A} & = & 2 \pi r \sin \theta ds = 
	2 \pi r \sin \theta \left( dr^2 + r^2 d \theta^2 \right)^{1/2} 
	\nonumber \\[1mm]
	& = & 2 \pi r^2 \sin \theta d \theta 
	\left( 1 + \frac{1}{r^2} \frac{dr}{d\theta} \right)^{1/2} 
\end{eqnarray}
or
\begin{equation}
d {\cal A} = 2 \pi r^2 d\mu \left( 1 + (1 - \mu^2) \frac{z'}{z}
\right)^{1/2}.
\end{equation}
Here we have introduced
\begin{equation}
\mu \equiv \cos \theta
\end{equation}
and 
\begin{equation}
z' \equiv \frac{dz}{d\mu}.
\end{equation}
Differentiating~(\ref{z}), the latter can be expressed as
\begin{equation}
\frac{z'}{z} = \frac{8}{27} \frac{z^3 \alpha^2 \mu}{2z - 3}.
\end{equation}
Putting the pieces together, we find that the luminosity is given by
\begin{equation}
L = 2 \int_0^1 2 \pi r^2 d \mu 
\left( 1 + (1 - \mu^2) \Big(\frac{z'}{z}\Big)^2 
	\right)^{1/2} F
\end{equation}
or
\begin{eqnarray} \label{lum}
\frac{L}{L_{\rm Edd}}
	& = & \int_0^1 d\mu
	\left( 1 + (1 - \mu^2) \Big(\frac{z'}{z}\Big)^2  \right)^{1/2} 
	\Big[ 1 - \\[1mm] 
	& & 2 \big(\frac{2}{3}\big)^3
	\alpha^2 z^3 (1 - \mu^2) + \big(\frac{2}{3}\big)^6
	\alpha^4 z^6 (1 - \mu^2) \Big]^{1/2} \nonumber
\end{eqnarray}
where $L_{\rm Edd}$ is the usual Eddington luminosity 
\begin{equation} \label{eddington}
L_{\rm Edd} = \frac{4 \pi M}{\kappa}.
\end{equation}
It proves most convenient to evaluate eq.~(\ref{lum}) numerically.  In
Figure~2, we plot the resulting luminosity $L$ as a function of the
spin parameter $\alpha$.  Obviously, for nonrotating stars with
$\alpha = 0$ we recover $L = L_{\rm Edd}$.  For maximally rotating
stars, however, the luminosity is reduced by about 36\%:
\begin{equation}
L_{\rm shedd} = 0.639 L_{\rm Edd}.
\end{equation}
Accordingly, adopting the Eddington luminosity for a supermassive star
that evolves along the mass shedding limit would {\em underestimate}
its lifetime by about 36\%.

\section{DISCUSSION}

We find that the luminosity of a SMS rotating at break-up velocity is
reduced by about 36\% from the luminosity of a nonrotating SMS of the
same mass.

It is difficult to compare this result with previous numerical
calculations of massive, rotating stars, which are summarized in Table
12.1 in Tassoul (1978, compare the discussion in Kippenhahn 1977).  No
calculations seem to have been performed for stellar masses greater
than $62.7 M_{\odot}$.  The luminosities of these $62.7 M_{\odot}$
models at break-up velocity are indeed reduced below the nonrotating,
spherical luminosities, but only by $~7\%$, much less than what we
find.  However, the physical conditions in $62.7 M_{\odot}$ stars are
very different from those in SMSs and do not satisfy our assumptions
(see Section 2).  For example, at these moderate masses, the stars are
not dominated by radiation pressure; according to eq.~(\ref{beta}),
$\beta \approx 1$ for these stars and $\beta$ varies with both the
location in the star and the orbital velocity\footnote{Note that if
$\beta$ were strictly independent of position and spin rate, the right
hand sides of eqs.~(\ref{f_eddington}) and~(\ref{eddington}) would be
reduced by a constant factor of $({\beta} +1)^{-1}$, and the
luminosity of a rotating star (eq.~\ref{lum}) would be still be
reduced by the same amount below the luminosity of its nonrotating
counterpart.}.  Also, the total opacity, at least close to the surface
of the moderate mass stars considered previously, is no longer
dominated by electron scattering and contains nonnegligible
contributions from bound-bound and bound-free absorption.  These
contributions introduce a dependence on density, so that our
assumption 4 no longer holds.  More specifically, spinning up the star
may decrease the density in the envelope, therefore decrease the
opacity, and hence increase the luminosity.  This would partly
compensate for the decrease of the luminosity due to rotation, and
will decrease the effect that we find for true SMSs.  Finally, we note
that moderate mass stars are not fully convective.

Our result is important because the luminosity determines the
timescale of the evolution and hence the lifetime of SMSs, which are
believed to evolve along a mass-shedding sequence.  This lifetime has
been used in several calculations of SMS evolution.  We adopt our new
result in Paper II, where we analyze the secular evolution of SMSs up
to the onset on radial instability.

\acknowledgments

This paper was supported in part by NSF Grants AST 96-18524 and PHY
99-02833 and NASA Grant NAG5-7152 to the University of Illinois at
Urbana-Champaign.

\end{document}